# The BIVEE Project: an overview of methodology and tools

EU needs an effective exit strategy from crisis, with a special attention to SMEs that represent the 99% of the enterprises active in the European production system. To this end innovation appears to be a key factor to relaunch the EU industrial system. The BIVEE project proceeded for almost 4 years to develop a rich framework, i.e., a methodology and a cloud based software environment, that includes business principles, models, and best practices, plus a number of advanced software services, to support and promote production improvement and business innovation in virtual enterprise environments (essentially, enterprise networks.)

**1.3.1. Framing**

European industrial system needs to change pace to relaunch its competitiveness, after a long downturn. To this end, new strategies need to include a renewed attention to continuous production improvement and, in a seamlessly way, business innovation. Both are essential for business survival in highly competitive markets where it is increasingly difficult to differentiate products and services. Rethinking the European value production system in a way capable of integrating different ideas, people, cultures, is the only way for Europe to compete with the impetuous progression of the BRICs (the rampant Brazil, Russia, India, China) economies.

Innovation is a term that represents a complex domain, requiring both domain expertise and a large amount of knowledge. Firstly domain knowledge (on the specific industry sector), but also knowledge on technology, business models, finances, markets, etc. Carrying out an innovation project is not an easily job for a single enterprise, then it is really a challenge for a network of enterprises, having the problem of achieving the necessary coordination and synergy in a distributed, multi-polar decisional and operational structure. We will refer to such networked enterprise structure as a Virtual enterprise (VE). The objective of the BIVEE project has been to building methodological and technological solutions to improve the value creation

---

Chapter written by *Michele Missikoff and Pierluigi Assogna*

capability of European SMEs, leveraging on constant production improvement and continuous business innovation.

In conceiving the project, we were aware that there are a number of barriers that EU SMEs still face and which prevent them from systematically adopting the emerging paradigms of collaborative business innovation. Among these barriers, we may cite the high fragmentation of the SMEs industrial fabric, the inadequate organizational and technological culture, the limited resources (such as financial, management expertise and competencies) available for SMEs, the lack of systematic connections with research institutions, and often their naïve and unfocused approach to innovation. However, there are a number of new business models and ICT solutions for supporting and fostering innovation in SMEs, such as Open Innovation, crowdsourcing, new forms of social networking, that we decided to acquire and transform into usable solutions for the European SMEs.

In BIVEE we assumed a broad, holistic approach to business innovation, tackling the methods, concepts, ideas, inventions, artefacts aimed at renovating in a sustainable manner the way enterprises conduct their production processes, considering that all is connected, at the business, technical, economical or social level. This is in line with the policy of the European Commission (see the Innovation Union Flagship initative[2]), and in particular the strategy aimed at portraying the next generation of innovation-driven enterprises: only a synergic combination of new technologies, new business models capable of supporting and respecting our social and political foundations, could succeed in making the Euroepan enterprise system evolving and pushing forward the entire socio-economic system.

Innovation largely remains a people-centric, 'brain intensive' activity. In this perspective BIVEE proceeded in building a distributed, collaborative, knowledge-intensive framework, where not only emerging ICT solutions but also innovative business models, novel management methods, and cooperative working styles have been integrated to the benefit of interoperable virtual enterprises.

The methodological framework of BIVEE is based on two interlaced spaces: Value Production and Business Innovation space. These are the concrete spaces, where resources and people cooperate to achieve the business goals, but at the same time they correspond to two knowledge spaces where the reality, inside and outside of the enterprise, is extensively represented in digital form. According to the BIVEE approach, such a rich knowledge base is used in the value production space (VPS) activities to support the monitoring and improvement of highly distributed production processes, and in the business innovation space (BIS) activities, to support the continuous innovation in a VE.

---

[2] http://ec.europa.eu/research/innovation-union/index_en.cfm

On the technological side, the logical architecture of the BIVEE Software Environment has been structured according to the two mentioned virtual spaces (VPS and BIS). Each of which is managed by a software platform: Mission Control Room (MCR) and Virtual Innovation Factory (VIF), respectively. Furthermore, to manage the VE knowledge supporting the activities of the two mentioned spaces, a platform for the management of a Production and Innovation Knowledge Repository (PIKR) has been developed. The three platforms stand on top of the Raw Data Management and Service systems that interface the local (to the individual SMEs) information systems, providing the necessary data interoperability and service integration for such diverse information sources. Among the offered services, we may list the support to a rich gamut of business activities: from production monitoring to market watch, from innovative think-tank support to technology watch, appraisal and evaluation of production and innovation activities, until risk assessment and prevention. An important position is given to innovation monitoring based on a number of Key Performance Indicators (KPIs) capable of tracking the progress of an innovation project and, subsequently, assessing the effects of the adopted improvement and innovation solutions. The four main platforms have been based on the Open Software Architecture paradigm and advanced meshup techniques, by extensively reusing resources available in the FLOSS world and leading edge solutions provided by BIVEE partners.

To validate the BIVEE Environment, the project addressed two trial cases: the first trial concerned the area of distributed co-design, having innovation as a core focus (in Loccioni - General Impianti); the second trial took place on a traditional production/logistic chain (AIDIMA). The involved industry sectors have been carefully selected to validate BIVEE in very different cases: in fact the latter concerns the 'mature' wood and furniture industry, while the former is positioned in the hi-tech sector of robotic and automatic measurement equipments. To achieve a thorough assessment of the BIVEE Framework, the trial cases have been organised in two main phases: Phase 1, the First Monitoring Campaign, where the activities of the two VEs have been monitored in their 'as is' practices, assessing their production and innovation performances before introducing the BIVEE platform; then, after the introduction of BIVEE, in Phase 2, we carried out the Second Monitoring Campaign, appraising the production and innovation performances once the BIVEE solutions have been adopted. Confronting the results collected in the two monitoring campaign has been a very useful activities that yielded to important indications for the implementation of the final version of the BIVEE Framework, released at the end of the project.

**1.3.2. The mission of BIVEE**

The mission of BIVEE is to provide advanced solutions to boost productivity and innovation capabilities of networked SMEs. To this end BIVEE developed an integrated framework, i.e., a coordinated set of business methods, enterprise models and ICT solutions, to support adaptive, distributed, interoperable business ecosystems (in our case also referred to as Virtual Enterprise Environments: VEE) in pursuing continuous, distributed optimization and innovation practices.

Production optimization and innovation need to be addressed in a synergic fashion. About the former, the main focus of BIVEE is on manufacturing that is mainly based on well defined, deterministic value production processes where tangible and intangible assets are managed in order to achieve some business objectives, in a cost effective way. Here automatic equipments, such as robots, measuring sensors, various actuators, play nowadays a very important role. The BIVEE solutions concern the services aimed at monitoring the distributed production activities and intervene to correct deviations to the planned programs and/or to introduce all possible improvements. In these cases the interventions do not structurally modify the enterprise organization or production maps (e.g., according to the *Kaizen* approach).

Innovation concerns the capacity to conceive and introduce a marked discontinuity, such as the introduction of a new product on the market or a new production process, that will impact on various enterprise dimensions, such as human resources, production means and methods, organization and finances, marketing strategies, etc. Innovation is a human-centric activity, often non deterministic or serendipitous, in most cases starting with imprecise objectives, presenting a number of criticalities in the planning, forecasting and scheduling the activities, and in the managing the risk. Here BIVEE proposes a specific set of services.

The lifecycle of an innovation project has three basic phases: the *inception*, the *evolution*, the *conclusion*. The inception typically starts according to five (not orthogonal) basic approaches: (i) *push-mode and technology driven*, when the innovation is generated on the supply side (e.g., by the appearance of a new technology); (ii) *pull-mode and demand driven*, when the innovation is motivated by a user need (i.e., by the future adopters); (iii) *co-creation*, when all the stakeholders cooperate together to generate product or process innovation (i.e., open, collaborative approach); (iv) *endogenous*, when ideas come from within the enterprise of the ecosystem; (v) *exogenous*, when ideas come from the rest of the world. We believe that the case (iii) is the most effective inception, although the hardest to achieve. In fact, innovative co-creation requires that different people, with different cultures, skills, and needs (and, sometimes, conflicting objectives) share a cooperation space spanning across the whole product lifecycle, from R&D, to design, to production, until after-sales, and across the enterprise boundaries. When stakeholders belong to different realities (different enterprises, but also different roles, different geographical areas, including also external research organization), often remotely interacting, with different levels of engagement and timing, then the added value is more important. A situation that is typical of Internet-based social networking.

The central phase of an innovation project is where ideas are consolidated and progressively elaborated, pushing them forward until an applicable innovative solution is achieved. This is the phase that will be elaborated in detail in this book, with a characterising point of view, concerning innovation projects carried out in a distributed, collaborative open organizations' network. Then the book addresses also the final phase of innovation and its interactions with the production space; in case of successful innovation the outcome will be transferred to production and, eventually, to the market. But also the inverse flow from production to innovation plays a crucial role, carrying over feedbacks on the adopted innovations, providing also stimuli for starting new innovation projects.

**1.3.4. Business ecosystems and virtual enterprises**

Business ecosystem and virtual enterprise are two central notions in BIVEE. The notion of a virtual enterprise (VE) refers to an organizational and business model where different production organizations (e.g., enterprises, production units) join together with a predefined objective (e.g., achieving a given production or business innovation) and share skills and competencies in order to attain a specific result (e.g., product, service, or a target market share) [AAG 99], [Byunghak, 2001]. Once the business objectives and the production plans are defined, a VE creates a Value Production Space (VPS), i.e., a complex networked business environment where the value creation activities take place. A VPS is organised in different layers (e.g. operational, resources, management layer, etc.), where different resources are allocated (e.g. humans, services, machines) and different activities take place respecting different constraints and factors (e.g., production capacity, budget constraints, laws, deadlines, etc.)

The most important characteristic involved in the selection of a production unit (PU) to participate in a VE is its capability. This characteristic reflects not only the availability of the required technology and relevant skill and experience, but also the reliability in terms of its records in respecting commitments, the capacity of producing the required volumes, the flexibility in relation to program changes, and so on. All this information is maintained in the business ecosystem repository (part of PIKR) and is constantly updated.

The coordination and collaboration of SMEs, or production units, in a single VE may raise specific issues, stemming from the fact that the different SMEs may significantly diverge in terms of organization model, management style, decision making methods, information transparency (but also 'hidden agendas'). Furthermore, different PUs and, more specifically, different production phases may be supported by different information systems (process models, data types, etc.) This diversity may cause that the information connected to outputs (products/services) released in one production phase is not fully (at least in an easy way) compatible with the input expected by PUs operating in the successive phase. To cope with this problem, BIVEE adopted a semantic interoperability solution, based on a central repository (including a set of ontologies) for the semantic annotation of the various resources by using a unified reference vocabulary and knowledge-based.

Furthermore, innovation requires continuous decision making, often a to be made in a short time. Since there is not a unitary center of command and there is a shared responsibility among the partners of the VE, there is the need to organize the VE information to support distributed decision making according to non conventional methods. This is achieved primarily by providing solid, reliable fact and knowledge base: a prerequisite to achieve informed decision making. The idea to measure, monitor and control innovation activities by using traditional time-cost-quality performance indicators is in general not suitable and in some cases could even be counterproductive, risking to seriously jeopardise an innovation project.

A second issue is a clear separation between local and global decision making. BIVEE provides the required functionality to overcome this dichotomy, allowing a loose integration (i.e., just to the extent needed by an effective cooperation), where local modelling methods and working styles remain independent, but at the same time offering the solution to provide a global view of such a fragmented picture. Effective global coordination with marked local autonomy present also the advantage of lowering the cost for a new enterprise to join a VE.

A Business Ecosystem (BES) [PEL 04] is a 'protected' space populated by enterprises willing to stay in touch in the perspective to get together to set up a VE. To be accepted in a BES an enterprise needs to satisfy a number of criteria and to respect some common rules, once included in the community. Criteria and rules are freely decided in a democratic way by the community itself. Entering in the BES, can happen by invitation or application. In both cases the enterprise must provide a substantial information, including competencies, skill, capabilities, previous experiences, that composes the enterprise profile. Such profiles are freely accessible by the other members of the BES, and automatically filtered by dedicated BIVEE services that support partner search during the composition of a new VE. The BES rules define also the exit mechanism for enterprises intending to leave, including 'non disclosure' obligations regarding the other partners information acquired while participating in the BES. Among the various advantages offered by the participation in a BES, there is also the collective procurement opportunity.

In a BES, enterprises and stakeholders operate in a participatory area (including both the Production and Innovation Space) organized according to three concentric circles, as depicted in Figure 1.3.1. The *inner circle* includes the PUs that participate in the VE or VIF (several VE/VIF can be active in a BES at a given time). The *middle circle* represents the trusted business ecosystem (BES) that gathers all the enterprises that have been recognized and qualified, and are therefore able to guarantee the required levels of quality, performances, trust and security. The *outer circle* represents the 'open sea', populated by all possible business players, potentially interesting for (and interested to) the ecosystem, but also player who compete in the same market and are potentially distrustful and hostile. The three circles offer different levels of openness and protection.

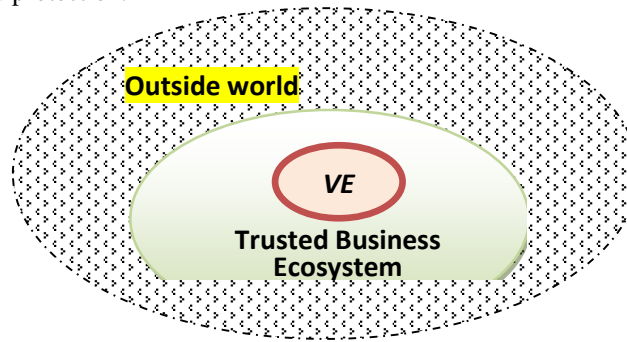

**Figure 1.3.0.** *A Participatory Space*

The key events of the BES lifecycle are listed below. BIVEE provides services to manage the following four basic events.
1. Evolution of the Business Ecosystem, when a new enterprise joins the BES or an existing one updates its capabilities or competences;
2. Exit of an enterprise from the BES, either for its explicit choice or for the progressive failing of the required membership criteria
3. Start of a new VE, in response to a new market opportunity;
4. Termination of a VE having reached its business objectives, or for the failing to fulfil the conditions required by the BES charter;

As anticipated, the BIVEE philosophy is based on two interleaved spaces where value production and business innovation take place. In the next two sections we will introduce the two mentioned space.

**1.3.5. Value Production Space**

The advent of the Internet has already offered new opportunities for enterprises to improve their business and production models, one of the key impact is represented by the increasing flexibility and adaptability of production carried out by a networked enterprise. Distributed, networked production models are difficult to design, implement, deploy, and manage in an optimal way. The 'traditional' approaches to enterprise optimisation are not suited and new solutions are sought to manage networked production paradigms, capable of adopting the power of the Internet.

In BIVEE we introduced the notion of a value production space (VPS), as a digital virtual reality aimed at modelling and representing a complex, distributed reality where a virtual enterprise operates. In a value production space we have production units (Pus, sketchily represented by rounded boxes in the Figure 1.3.1) that can be independent SMEs or organizational units in a single enterprise (we don't need to distinguish at this level). A PU can be of four different sorts: manufacturing, assembly, service, logistics. A production map is represented by a graph, with nodes that represent production units connected by flow arcs. In a production map, that includes also other complementary infrastructures such as storage warehouses and other necessary services, there are branches with alternative or parallel paths (a path is a linear subgraph). The primary role of a value production map is to represent the flow of goods and services, but also other flows, such as the financial and information flows. A value production process evolves over the production map, traversing a number of (pre-) defined paths, links and units.

In our view, a value production space (and the VE operating therein) is not a closed territory: it is open for (impromptu) contributions from the rest of the BES and, in different forms, from the external world; at the same time it is sufficiently protected to ensure the required levels of trust and security to the members operating within such a space (as exemplified by the dotted border line of the production space in Fig. 1.3.1 that represents a VE with its PUs). BIVEE provides the services and the methods to model the VPS, the VE, and to monitor the activities that take place along the production paths.

The Mission Control Room implements the services that BIVEE offers for managing the value production processes taking place in the VPS. In particular, it enables the users to:
- Create and maintain the Business Ecosystem, that is the "incubator" of new VEs;
- Support the creation of a VE, helping the discovery and selection of partners, and possible evolutions of its composition;
- Define and maintain the production map of each VE;

- Monitor the distributed production activities, by gathering feed-back from each PU in the VE
- Periodically update the production plan distributing the updated tasks and workload to the interested PUs in the VE;

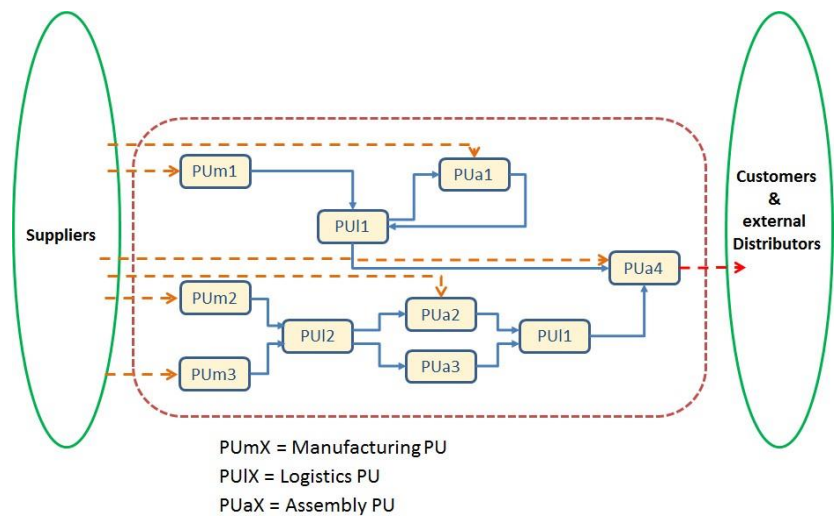

**Fig. 1.3.1**. *A Value Production Map*

As seen, while guarantees that the daily routine is constantly controlled for maximum adherence to programs, has the objective of improving current operations. This objective is supported by the production monitoring based on a number of ad-hoc Key Performance Indicators, tailored to each specific VE.

In setting up a new VE, BIVEE supports the development of two main documents:
- **Business Plan**, that is focused on the creation of the VE, starting from the production objectives, the market to be served, and the production resources involved; it indicates also the expected revenues and costs. One of the main supports provided is the screening of the capabilities of the candidate Production Units, in term of type of production, skills, acceptable volumes, time responsiveness, in order to configure the VE.
- **Master Production Plan**. This document comes afterword and concerns a given delivery plan; it is compiled after having verified the production objectives, with quantities and times, then the feasibility, by the configured VE, in terms of available resources, cost, schedule, and so on.

Once the VE has been set up, the platform provides support to the continuous production monitoring and management activities. These have been formalized through a set of process templates (see Chapter 2) focusing on the production process as seen at VE level, where each component PU is seen as a work centre of a traditional ERP (Enterprise Resource Planning system). The support to the local production management and control system of each single PU is not considered part of the BIVEE mission.

The Value Production Space processes are described according to the SCOR Framework[3] that introduces precise definitions of the functionalities mentioned above. Summarising, the primary production processes of the VE are (according to SCOR): *Value Chain Plan, Production Plan, Production Source, Production Build, Production Deliver*. These processes are applied to the VE in its entirety, but each PU will take care of specific tasks and will organise its own internal processes accordingly. The VPS processes, seen at VE level, are represented by networks of macro activities performed by the involved PUs. For example *Production Assembly*, an activity that can be found in the *Production Build* process, does not detail the specific sequence of operations in the local PU, but sees it as a 'black box', considering in input the consumption of the required components, and in output the staging of complete assemblies, according to the Master Production Plan.

Plans that are distributed to the involved PUs can require some later negotiation in case of unpredicted events. For this reason, advancement feedback is harvested from the PUs, in the form of KPI measures, to create a picture of the progress in the production processes. In case that a PUs is not able to send the required information, for instance if its enterprise system is down, the BIVEE platform provides a Web-based user service for manual data collection.

*Production knowledge support*

Focusing on the value production activities of a VE, a central role is played by the Production and Innovation Knowledge Repository (PIKR). The objective of this tool is to maintain a constantly evolving knowledge base, aligned with the state of play in the production field, extended along various dimensions:
1. Knowledge sources: the PIKR contains crowd-sourcing mechanisms, in order to harvest knowledge by an extended audience of business, market, technology experts, that may significantly help in production improvement.

---

[3] Supply-Chain Operations Reference (SCOR) model: http://www.supply-chain.org/

2. Organizational memory: the history of all improvements and innovation exercises, whether successful or not, is maintained and can be retrieved in order to evaluate results and lesson learned.
3. Similarity of topics in different production activities: production improvement often stems from addressing the issues of a product or a process with solutions used in different domains. The availability of "knowledge nuggets" coming from more or less contiguous territories can trigger interesting searches of new materials, operations, applications, etc.

*Business Ecosystem*

To be considered as candidate for participating in a VE, an enterprise has to be screened on the base of a number of characteristics (reliability, technology level, production capacity, etc.) Enterprise profiles, maintained in the PIKR and adjusted to the evolution of the state-of-play, are analysed in the context of the VE interests.

New members of the BES considered for membership after they explicitly apply (push mode) or by invitation (pull mode). The scouting of possible new entries in the BES is supported by the Observatory function; when a new VE needs to be created and none of existing BES members is eligible, the VE manager can activate a BIVEE search to screen enterprises existing in the "outer world" that exhibit interesting characteristics. Out of the list provided by BIVEE semantic crawler, the VE manager can select candidates that can be contacted and invited to join the ecosystem first, and the VE afterwards.

*Product and/or Process Improvement*

VE changes aimed at achieving the planned improvements need to be implemented in a way that minimise the impact on BAU. Therefore, it is important to maintain a clear separation between current production programs and executions, on the one hand, and feasibility of changes, assessing their effectiveness, on the other.

The triggering of improvement requests, management of the changes, analysis of results, etc. are supported by the functions of the MCR, while the actual test of the implemented improvement, and the relevant feedback, is managed by the VPS monitoring services: the main knowledge objects involved in these services are KPIs, elaborated below.

*Ongoing VE production*

The SCOR modeling frame is the base of the MCR services that are grouped into the following components: *Modeler*, aimed at the VE Setup, including the modelling of production maps in the VPS; *Assistant*, that supports the rollout of the Master Production Plan, with its production processes; *Monitoring*, that manages the feedback gathering from the operational field, based on the KPIs coming from the PUs. The knowledge objects involved in this case are on one side the Production schedules for the PUs, and on the other the production advancement, reflected by the KPI measures, harvested from the PUs.

Figure 1.3.2 shows the relation of MCR with PIKR and the data exchanged with the PUs. Please note that such a data exchange actually takes place by using the Raw Data Management services.

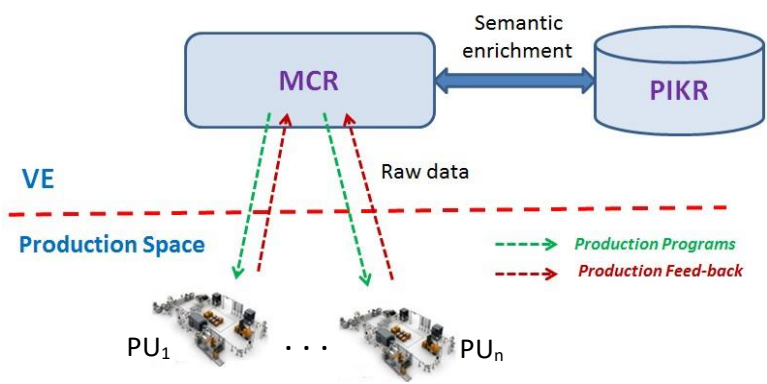

**Figure 1.3.2.** *MCR information exchange*

### 1.3.6. A Participatory Space for business innovation

One of the primary goals of BIVEE has been to address innovation with an 'industrial' philosophy. In this perspective, innovation is seen as an intangible 'knowledge artefact' to be built starting from 'raw knowledge', suitably selecting and connecting the required 'knowledge parts', then proceeding to create the 'missing parts' needed to realise the final artefact, i.e., the body of knowledge that represents the sought innovation. The conception of such an engineering approach to innovation has been particularly challenging, since we had a narrow path to go, trying to push the proposed methodology towards a more rigorous and systematic approach, without jeopardizing the creativity required by innovation. Furthermore, the innovation process is inherently less defined than a production process. In this frame we introduced the notion of a Virtual Innovation Factory, based on the following elements:

- Human experts, who are considered, with their intuitions and creativity, the primary engine of innovation
- A set of guidelines, intended to be used by the innovation teams in a flexible and discretional fashion
- A knowledge repository, holding all (possible) information that can be useful during the innovation project; it holds pre-existing knowledge, as well as the new knowledge produced (and acquired) during the work.
- A virtual collaboration space, where the distributed innovation teams can exchange ideas, questions, comments, answers. The collaboration, that can be synchronous or asynchronous, is always supported by an ubiquitous, easy access interface to the knowledge repository.
- Monitoring and assessment services, supported by a platform for collecting facts and evidences from the various innovation teams, feeding then the KPI Management System that will constantly provide the state of play of the project, based on quantitative elements.

The VIF eventually produces innovation artefacts that are passed to the VE and then applied to the VPS to concretely implement the innovative solutions. In rolling out the innovative solutions, the VE needs to carefully understand the cascading effects of the proposed changes. Such changes generally require the re-alignment of the VE over several dimensions and enterprise areas, beyond the one initially targeted by the sought innovation (here Change Management is centrally involved, not directly addressed by BIVEE since it is a well elaborated discipline, very rich in proposals and scientific results [PAT 08]).

In summary, Business Innovation is a designed, managed transformation of some aspects of an enterprise aimed at a substantial improvement of:
- the quality of delivery products (goods, services) and the customer satisfaction,
- production processes and workers satisfaction
- cost reduction and/or revenue increase
- sustainability of the production (i.e., implementing transformations respectful of working conditions, social context, environment, ...)

An innovation project typically starts with a (more or less) defined objective that is positioned in one focus dimension (e.g., introducing a new product in the catalogue), then it is necessary to identify all the other dimensions that are impacted by the specific innovation. Often, such an impact is overlooked and that implies a high probability of failure. The primary dimension that always is involved concerns enterprise processes, but to a certain extent all the other dimensions are affected..

The business innovation space is organised in a different way with respect to a value production space. The latter typically transforms raw material into finished products (or elementary services into complex services); while in the innovation space we take existing production processes aiming at producing new processes. So, innovation primarily operates on the enterprise itself, that is the object of the transformation. Furthermore, an innovation process is inherently different from a production process, since the former is essentially ill-defined, typically built ad hoc. Even if it is based on existing guidelines and make use of pre-existing practices (i.e., guidelines, sub-processes), large space is left (hopefully) to creativity, intuition, 'lateral thinking' that hardly support to be encapsulated in systematic, repetitive tasks. For this reason, we prefer to avoid the term 'process' (implying a well defined and organised set of activities repeated in time) when we talk about innovation, being 'project' more suited.

An innovation project sees the initial involvement of creative units, followed by specialised units addressing specific issues, such as design, engineering, financial and market ones. All the units cooperate and interact by means of (generally distributed) communication platforms supporting the continuous exchange of ideas, comments and suggestions, design artefacts, mock-ups, blueprints, but also quantitative data and simulation outputs.

Similarly to what introduced for a VPS, a business innovation space is characterised by an *innovation map*: a connected graph with nodes representing activities (and the teams enacting them) and arcs representing the flow of knowledge. As opposed to a production map, an innovation map is not fully defined when the activities begin, it is dynamically specified while it is traversed along the time, and new steps are dynamically identified, on the basis of the achieved results and the encountered problems. The traversed paths are aimed at progressively increasing the knowledge necessary to achieve the sought innovation. Such a progression typically starts from the existing state of play, a consolidated set of products and production process, gradually acquiring new knowledge until the innovative solutions are fully specified. In the most radical cases, the innovation project may require the extinction of (part of) what exists, to be replaced by radically new solutions[4]. In parallel, we need to consider the risk that an innovation project may fail, therefore it is necessary to show, with reliable forecasting methods, that the undertaken direction has good chances to terminate delivering the expected results.

---

[4] According to Joseph Schumpeter's Creative Destruction: Schumpeter, Joseph A. (1994) [1942]. Capitalism, Socialism and Democracy. London: Routledge. ISBN 978-0-415-10762-4.

*Innovation management*

At the core of the Business Innovation Space is the concept of innovation management. Innovation management is the process of managing ideas through the stages of the innovation cycle until they reach an industrial maturity and eventually the market [HAM 07]. The innovation cycle describes the activities involved in taking an innovative idea, elaborating it to progressively achieve industrial strength products or services. [ATT 15]

Generally the innovation cycle has been represented as a process flow taking place within a single enterprise, where the interrelations with the external environment are mostly marginal and external collaborations mainly provide punctual contributions. Such a form of 'closed innovation' has been primarily motivated by the need to protect ideas and industrial innovation plans.

Starting from 2003 the paradigm of closed innovation has been debated and contrasted with the emerging paradigm of Open Innovation, firstly introduced by Henry Chesbrough [CHE 03]. Open Innovation has nowadays taken a permanent place in the business and organizational world. In addition, the advent of new ICT solutions and the predominance of Internet based services have made the enterprise environment progressively more open and participatory. One of the main issues of open, collaborative innovation concerns the protection of ideas and the intellectual property rights (IPR): there is the need of finding a virtuous balance between the need to disclose ideas and knowledge, to facilitate (even unexpected) external contributions, and to protect the key knowledge assets, to avoid that even partial results can be acquired and used by competitors.

In BIVEE we proposed the Business Innovation Reference Framework (BIRF) to organise the BIS according a rigorous method while preserving the necessary flexibility. The BIRF is based on three main pillars: (i) the BIVEE Innovation Waves, (ii) the document based collaborative knowledge management, and (iii) the monitoring strategy, based on carefully selected KPIs. All the three pillars are implemented by the collaboration space of the Virtual Innovation Factory (see below).

**1.3.6.** *BIVEE Innovation Waves*

An innovation project, typically triggered by the 5 anticipated cases (see Sect. 1.3.2), starts by carefully considering new technological and market opportunities, having also in mind specific client needs. This is achieved with an exploration and scouting phase, including the access to pertinent knowledge available at well renown research and innovation centres (of partners, agencies, universities, etc.) Then, according to BIVEE, an innovation project is organised following the 4 BIRF Innovation Waves: Creativity, Feasibility, Prototyping, and Engineering, as described below.

**Creativity Wave**: This first wave starts with an innovation idea or a problem to be solved, providing a first description. Then, a team is established that starts to elaborate the initial idea, searching for similar ideas and previous experiences (including past failures) related to it. If the idea is promising, a Virtual Innovation Factory is created, with one or more teams with resources belonging to different real enterprises.

**Feasibility Wave**: This wave starts when the initial idea is sufficiently specified, with its structure, functions, scope, and intended impact, and approved by the corresponding manager. Then it proceeds elaborating the business plan, market analysis (including competitors, customer profiling) and the feasibility documents (e.g., technical, financial, industrial, market feasibility), with risk analysis and patent possibilities. The output of this wave, that includes a preliminary design, is checked by the top management that, in case of positive evaluation, allocates the required budget.

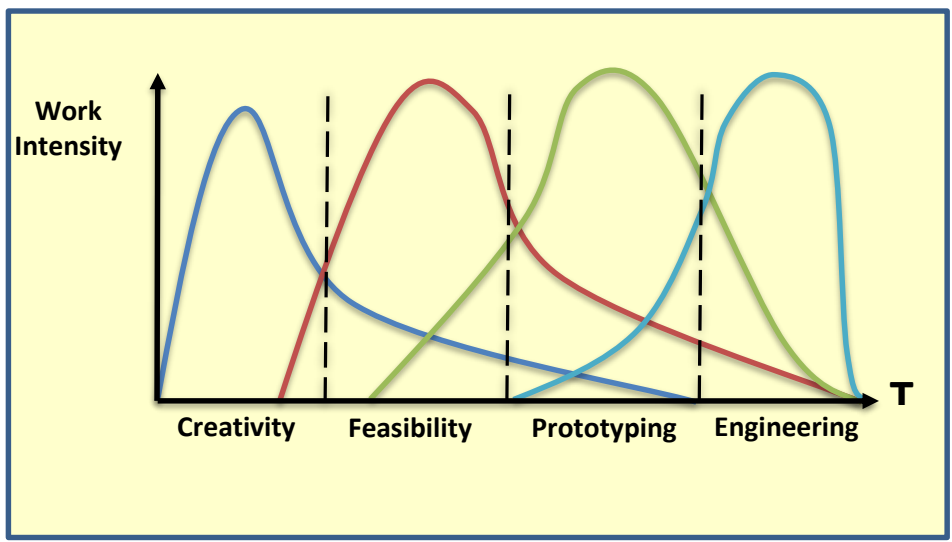

**Figure 1.3.3.** *The four innovation waves of BIVEE*

**Prototyping Wave**: This wave starts elaborating the detailed design and proceeds with the first implementation of the initial idea. Simulations and extensive testing (in the lab, as well as in the field) are carried out. This is the wave in which the innovation is concretely projected into the real world for the first time, allowing the initial idea to be confronted with implementation and practical issues.

**Engineering Wave**: This is the final wave that concerns the industrialization of the innovative solutions, with the documentation to be transmitted to the VE to start the rollout of the innovative solution in the value production space. Also the prospective budget, break-even point, and business model are defined, together with all the other documents that concur to form the knowledge asset necessary to activate the industrial production. Such knowledge includes documents like the Bill of Material (in case of a manufacturing product), production plans (indicating what to make or buy), delivery strategy and set-up instructions, testing and maintenance procedures.

We call them waves since they are logically sequenced in the time, but they are tightly interconnected and the start of a new wave does not imply that the previous one has been fully accomplished. Furthermore, there will be often the need to jump back and forth to complete a document or to correct it on the bases of later findings. For instance, during the prototyping wave there can be new findings that jeopardize the results of a previous financial feasibility study, requiring therefore to rethink some parts of the innovation under elaboration. The wave approach is sketchily depicted in Figure 1.3.3.

The innovation waves represent a powerful framework to guide innovators in achieving an innovation project. They are sufficiently powerful to be used for different kinds of innovation (e.g., product, process, market, etc.) and in different application domains (from automotive to tourism, from health to Government.) The guidelines associated to the waves indicate a number of document templates that, to be filled, require specific knowledge to be gathered. Waves are also associated to a number of carefully conceived Key Performance Indicators and a method to monitor and assess the progress of the work.

*The architecture of the VIF platform.*

The BIRF framework represents the rationale of the Virtual Innovation Factory (VIF), and in particular of the software platform that supports the innovation projects. The VIF platform has been conceived to support collective knowledge creation and management of innovation. To this end, the primary components are:

**Shared Semantic Whiteboard** (SSW): This is a knowledge sharing and visualization Web platform, where each member of the innovation team can post ideas, comments, issues, etc. The SSW can be remotely accessed, it displays elements that can be seen, edited, commented by all the members of the team (both in sync and async mode). Furthermore, the platform proactively searches and associates to SSW elements relevant knowledge items (e.g., documents) extracted from the Open Innovation Observatory and the PIKR (see below).

**Open Innovation Observatory** (OIO): This is a knowledge repository where the dedicated search engine of the VIF collects and organises the material extracted from various public sources, both on the web (the open section) and on the partners local repositories. Among the stored knowledge there is information about worldwide excellence centres, various projects and scientific results, relevant for the community. A careful and continuous observation of the evolution of the external world, managed by OIO, guarantees that BIVEE holds the information necessary to timely trigger suitable optimization and/or innovation actions.

**Collaborative Innovation Capability Maturity** (CICM): the BIVEE framework has also proposed a set of criteria and guidelines for networked SMEs (and enterprises in general) to (i) verify the innovation readiness of the enterprises and (ii) to identify a route that an enterprise can follow to progressively improve its innovation capability. The innovation CMM Model is inspired by the well known capability maturity model originally introduced for the Software Engineering by the SEI (Software Engineering Institute), but in addition to the 5 levels that concern the CMM of a single enterprise (Initial, Repeatable, Defined, Managed, Optimized) it introduces another dimension to cater for the virtual (networked) aspect of the VE. The latter dimension is in turn organised in 4 levels: Single organization, Network Awareness, Network Consent, Network Dedication. The CICM model has guided the implementation of a web-based tool for the self assessment of innovation maturity of a VE (http://innonetscore.de).

### 1.3.7 An integrated view of VPS and BIS

Business innovation and production improvement both have in common the fact that they aim at changing for better some aspects of the enterprise. As anticipated, a crucial problem is the identification of the key enterprise dimensions where such changes will take place. It is obvious that the enterprise dimensions are not independent one another and, in general, the first changes will trigger a sort of chain effects, with a more or less wide impact. In a broad view, the enterprise transformations can involve the following dimensions: *Product, Service, Process, Technology, Organization, Market, Strategy.* In BIVEE we decided to focus on the first four dimensions.

Today, the transformations are increasingly happening at a fast pace, with a trend towards a continuous improvement / innovation paradigm. Figure 1.3.4 reports a sketchy representation of two nested cycles, one concerning the optimization cycle and the other the innovation cycle.

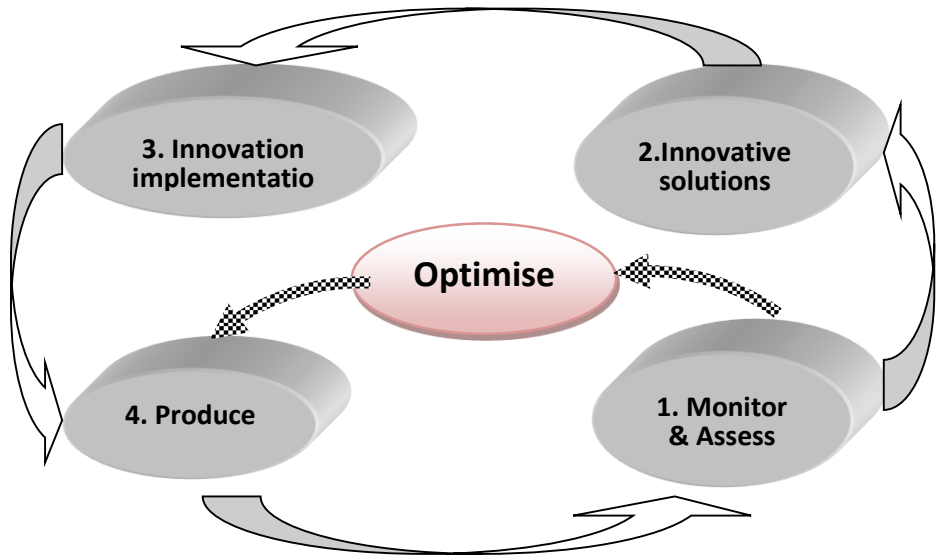

**Figure 1.3.4.** *The Optimization-Innovation Cycles*

According to the BIVEE approach, the optimization-innovation cycles are tightly interwoven, supported by the two application platforms that we have seen above: the Virtual Innovation Factory and the Mission Control Room. In fact, the two corresponding spaces, BIS and VPS, are also tightly interwoven, while maintaining respective roles, objectives, and characteristics that are inherently different. As anticipated, the idea is that an innovation map takes as input a 'consolidated' value production map and generates as a result a new, innovative, production map.

The Figure 1.3.5 sketchily represents such an integrated view, where the upper part shows the existing production map (PMx), i.e., the AS-IS situation that is going to be changed by the sought innovation; the central part symbolise the innovation space where an innovation solution is progressively elaborated; the lower part illustrates the innovated production map (PMx'), after the innovation has been fully implemented. The BIVEE platform essentially operates on this meta-space, adopting the most effective knowledge representation methods and notations.

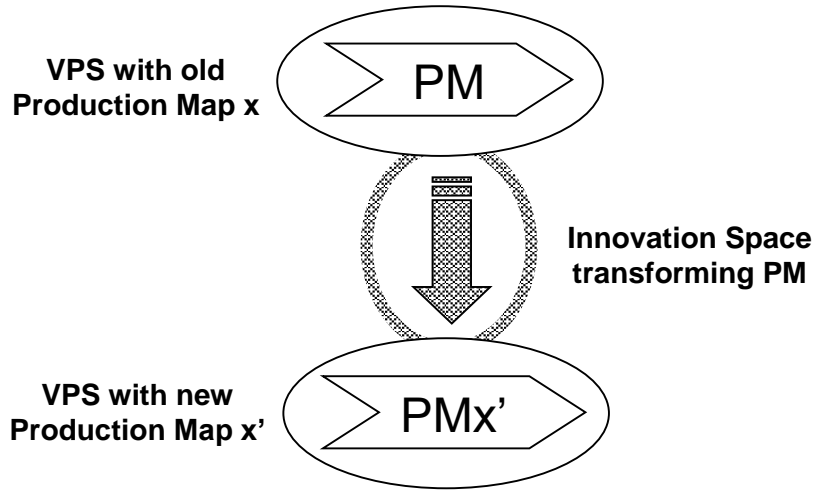

**Figure 1.3.5** Innovating a Production Map

And now the Figure 1.3.6 showing the interplay of the VIF and the MCR.

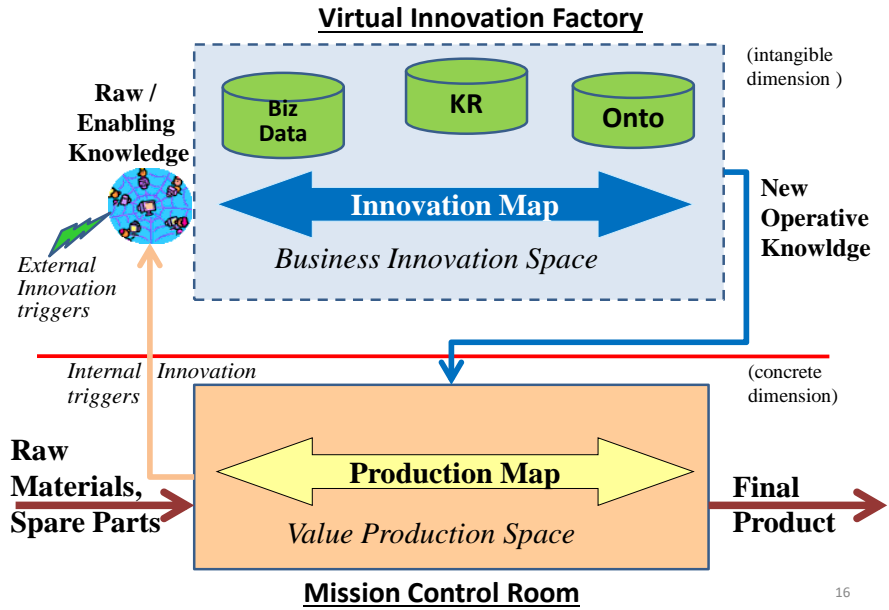

**Figure 1.3.6.** The interplay between VIF and MCR

**1.3.7. The Macro-architecture of the BIVEE Platform**

In the foreseeable future there will be a strong impact on organizational models where enterprise application and services will undergo a transformation of the traditional monolithic on-premise systems (such as ERPs) into heterogeneous collections of services, available from different providers on-demand and, e.g., on a pay-as-you-go basis. The traditional three layers architecture *data-application-presentation* is now implemented by services which allow for a boundless access to heterogeneous data (e.g., with a linked open data approach), for the accomplishment of enterprise operations (planning, forecasting, monitoring, management) and the implementation of new forms of mobile and fixed human-computer interaction (the so-called *Service Front Ends*). *Cloud Computing* is also pushing forward a comprehensive service paradigm to implement infrastructure-, platforms-, and software-as-a-Service (IaaS, PaaS, SaaS) solutions. Finally, some functions previously implemented in the application part of the enterprise system are now moving to the infrastructural part (see for instance the idea of an ISU: *Interoperability Service Utility*), becoming new kinds of network utilities. We envisage that Business Intelligence, and more generally Big Data Analytics, services will be among the key enablers for a virtuous growth of EU enterprises and will be among the value-added services that will characterize next generation enterprise applications capable of influencing new styles of fact-driven management.

In this frame, the macro-architecture of the BIVEE Platform is conceived to put the business user in the centre. In particular, the idea is that managers will progressively become 'coaches', active in monitoring and managing both production and innovation spaces with an open, participatory style. To this end, the organization of the knowledge repositories, the offered services and, overall, the user interfaces, will be conceived to create a comfortable, familiar environment for the (least technical) users. For this reason, particular attention have been placed to achieve advanced graphical user interfaces for the MCR and the VIF, where business people will collaborate in monitoring and managing enterprise entities (much in accordance with the principles indicated FInES Research Roadmap 2025[5]).

The BIVEE software environment, besides the platforms conceived for the end users, MCR and VIF introduced above, includes the following platforms. Please note that all the platforms and services introduced in this chapter will be addressed in the book by the chapters specifically dedicated to each of them.

**Production and Innovation Knowledge Repository** (PIKR): This repository contains the data and knowledge concerning production and innovation activities coming from the different partners of the VE and VIF, with the objective to provide a wide but 'controlled' knowledge sharing. It complements the OIO that, conversely, holds public 'external' knowledge. In particular, PIKR stores knowledge about all the teams working in the BES with their competences and experiences, the achievements of past projects, but also information of the VE production space evolution. PIKR is centrally based on the various ontologies: *DocOnto*, for the business document templates, *ProcOnto*, for business process schemes, and *KPIOnto*, for the various key performance indicators. The PIKR, that adopts particular security mechanisms to protect its content, is a virtual repository since the actual resources physically reside locally, at VE/VIF's partners sites: it hosts and manages ontology-based images of such resources.

---

[5] Future Internet Enterprise Systems Research Roadmap. EC Task Force Report.
http://cordis.europa.eu/fp7/ict/enet/documents/fines-research-roadmap-v30_en.pdf

**Production & Innovation Key Performance Indicators** (PI-KPI): one key role of BIVEE is the monitoring of the achievements of the PUs working in a distributed, collaborative fashion. To this end, the project proposes various key performance indicators, both for production (P-KPI) and for innovation (I-KPI). There is a rich literature proposing a large number of performance indicators, different for goals, domain of interest, degree of precision, and formalism, coming, e.g., from international and national standardization bodies, including reference models like the SCOR (Supply-Chain Operation Reference model[6]) and the VRM (Value Reference Model[7]). KPIs in BIVEE are evaluated starting from the information coming from the field operations and represent an invaluable asset on which informed decisions can be made. The relevant KPIs are organised in a hierarchical structure, as specific indicators can be grouped in a higher level ones: a specific tool, the KPI Modeller, enables business experts to easily create these structures. The Monitoring Framework (supported by the *KPIOnto* ontology) currently holds a total of 69 KPIs for the production space, 100 KPIs for the innovation space, and 29 KPIs concerning the sustainability context in both spaces.

**Raw Data Management and Service** (**RDM&S**) platform. This is the enabling platform that connects the various information sources, local to the different enterprises cooperating in the value production space or the business innovation space.

The RDM&S platform guarantees a constant surveillance of the ongoing activities, within the MCR and VIF, collecting and homogenising the key performance measures (KPM) that represent the factual evidences necessary to derive the key performance indicators (KPI).

---

[6] http://supply-chain.org/f/SCOR-Overview-Web.pdf
[7] www.value-chain.org/value-reference-model

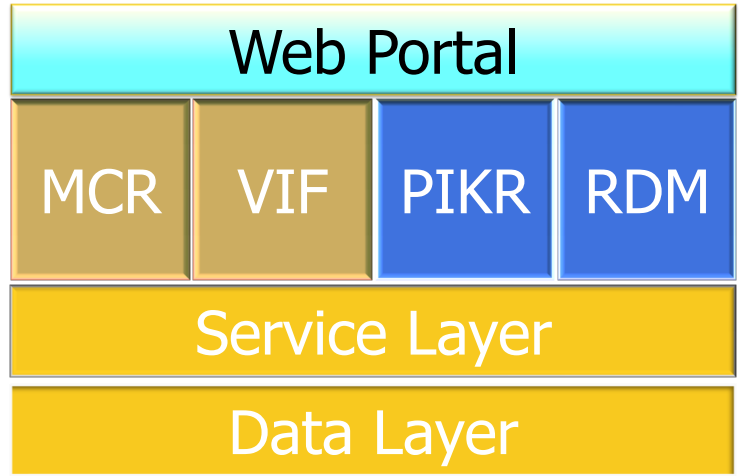

**Figure 1.3.8.** *The BIVEE Runtime Framework*

Figure 1.3.8 sketchily illustrates the global architecture of the BIVEE software environment. It highlights the web-based user interface on the top and the service-data layer on the bottom. In the middle the four software platforms that provide the end-user services (the two blocks on the left) and the knowledge-data services (the two blocks on the right.)

**1.3.8. Trial cases and Impact**

Impact creation has been one of the central goals of BIVEE and therefore an integral part of the project. To this end, two pilots have been carefully selected to provide a good coverage of different industrial cases; in fact, they are positioned at the two extremes of the technological scale: one belonging to the low-tech sector of wood and furniture and one to the hi-tech sector of robotics and automated measuring equipments. This strategic choice aimed at demonstrating that the value proposition of BIVEE can really have a wide scope. The second qualifying choice has been to carry out a comparative assessment in two distinct phases, launching two different monitoring campaigns. The First Monitoring Campaign (FMC) aimed at acquiring evidences about the performance of production and innovation activities in the 'as is' situations before the adoption of BIVEE. The First Monitoring Campaign, besides collecting quantitative and qualitative information from the operational fields, brought to the project the strategic advantage to involve end users as early as possible. Getting in contact with the future adopters of BIVEE give us the possibility to introduce them to a number of clues, while allowing us to gain a better understanding of a number of issues that were scarcely considered in the initial design of the software environment. Such issues, if underestimated in the early design phase, would have caused a significant delay in the later phases of the project. This has been an important lesson for the engineers and designers, understanding that a good expertise and a strong theoretical background need to be always confronted with the real needs of the users and the stakeholders.

After the FMC, we finalised the first BIVEE prototype, starting to progressively deploy it, in order to begin the Second Monitoring Campaign (SMC) aimed at gathering the information on the two trial cases observed in presence of the BIVEE platform. The objective has been that of confronting the evidences collected in the SMC with the corresponding evidences collected in the FMC. As usual, the reality fully reveals its complexity only when you start to practice it, any kind of theoretical model will be hardly able to seize it. Therefore, in the SMC we realised that a systematic contrast with the FMC data could be hardly achieved. The reality proved to be faster and slower than expected. Faster, since in a year many things have changed, even in the same industrial context (due to the international crisis, to split and merge of enterprises, etc.) and slower, since we realised that many enterprise phenomena, especially those connected to the structural changes, need time to take place. Therefore, a monitoring campaign of six month is too short to cover a significant time span, both in case of improved production processes and innovation projects. Nevertheless, we were able to collect several evidences showing the positive impact induced by the adoption of BIVEE. At a strategic level, the large majotity of users and stakeholders agreed that BIVEE is going to occupy an important area where there is very little offering and, at the same time, where there is a manifest need. Such a need will be steadily growing in the future, since even when the economy (hopefully) will restart, the call for more effective management of production improvement and innovation will increase. And solutions like BIVEE will be needed furthermore.

The main lesson learned concerned the limit of technology, especially if introduced in a working context where there is still a digital gap. There is a risk embedded in the idea that good technical solutions will 'automatically' improve to way people operate, make decisions, collaborate, produce value. We already knew it, but we had a further evidence that when we deal with socio-technical systems, such as the BIVEE Environment, the 'socio' component plays a central role. Therefore, when designing such a kind of systems, in parallel to the development of the technological solutions, it is necessary to consider, with a marked collaborative and participatory approach, also the new socio-organizational solutions, including educational initiatives for the involved people.

**1.3.9. Bibliography/References**